\documentclass[10pt, letterpaper, onecolumn, draftcls]{IEEEtran}
\AtBeginDvi{}
\usepackage[cmex10]{amsmath}
\allowdisplaybreaks[1]
\usepackage{amssymb,amsthm}
\usepackage{mathrsfs}
\usepackage{cite}
\usepackage[dvipdf]{graphicx}
\usepackage{bm}

\theoremstyle{definition}
\newtheorem{theorem}{Theorem}
\newtheorem{lemma}{Lemma}
\newtheorem{corollary}{Corollary}
\newtheorem{remark}{Remark}

\newtheorem{proposition}{Proposition}

\newcommand{\eqtri}{\triangleq}

\newcommand{\bX}{\bm{X}}

\newcommand{\cX}{\mathcal{X}}

\newcommand{\cA}{\mathcal{A}}
\newcommand{\cB}{\mathcal{B}}
\newcommand{\cC}{\mathcal{C}}

\newcommand{\cS}{\mathcal{S}}
\newcommand{\tcS}{\tilde{\cS}}
\newcommand{\bcS}{\overline{\cS}}
\newcommand{\cT}{\mathcal{T}}

\newcommand{\hQs}{\hat{Q}^*}
\newcommand{\tQl}{\tilde{Q}^{(\lambda)}}

\newcommand{\Ex}{\mathbb{E}}
\newcommand{\abs}[1]{\left\lvert#1\right\rvert}
\newcommand{\funcabs}[1]{\left\lVert#1\right\rVert}

\newcommand{\code}{\Phi}
\newcommand{\error}{\mathrm{P}_\mathrm{e}}

\title{On the Smooth R\'enyi Entropy and Variable-Length Source Coding Allowing Errors}
\author{Shigeaki Kuzuoka~\IEEEmembership{Member,~IEEE}
\thanks{The work of S.~Kuzuoka was supported in part by JSPS KAKENHI Grant Number 26820145.}
\thanks{S.~Kuzuoka is with the Faculty of Systems Engineering, Wakayama University, 930 Sakaedani, Wakayama, 640-8510 Japan, e-mail:kuzuoka@ieee.org.}
}

\begin{document}
\flushbottom
\maketitle

\begin{abstract} 
In this paper, we consider the problem of variable-length source coding allowing errors. 
The exponential moment of the codeword length is analyzed in the non-asymptotic regime and in the asymptotic regime.
Our results show that the smooth R\'enyi entropy characterizes the optimal exponential moment of the codeword length.
\end{abstract}


\section{Introduction}
Renato Renner and Stefan Wolf \cite{RennerWolf_isit04,RennerWolf05}
introduced a new information measure called the \emph{smooth R\'enyi
entropy}, which is a generalization of the R\'enyi entropy
\cite{Renyi61}.  They showed that two special cases of the smooth
R\'enyi entropy have clear operational meaning in the fixed-length
source coding problem and the intrinsic randomness problem: (i) the
smooth max R\'enyi entropy $H_0^\varepsilon$ characterizes the minimum
number of bits needed for with decoding error probability at most
$\varepsilon$, and (ii) the smooth min R\'enyi entropy
$H_\infty^\varepsilon$ characterizes the amount of uniform randomness
that can be extracted from a random variable.

As the notations indicate, the smooth max/min R\'enyi entropies
$H_0^\varepsilon$ and $H_\infty^\varepsilon$ are defined as limits of
the smooth R\'enyi entropy $H_\alpha^\varepsilon$ of order $\alpha$; see
Section \ref{sec:renyi_entropy} for details.  Hence it is natural to ask

\smallskip\noindent
\emph{
Does the smooth R\'enyi entropy $H_\alpha^\varepsilon$ of order $\alpha$ have operational meaning?
}
\smallskip

In this study, we answer this question by demonstrating that the smooth
R\'enyi entropy characterizes the optimal exponential moment of the
codeword length of variable-length source code allowing errors.
Our contributions in this paper are summarized as follows.  

\subsection{Contributions}
We consider $\varepsilon$-variable-length source coding problem, that
is, a variable-length source coding problem where decoding error is
allowed as long as it is smaller than or equal to the given value
$\varepsilon\geq 0$.  Usually, in this setting, the average codeword
length $\Ex[\ell(X)]$ is investigated; see, e.g., \cite{KogaYamamoto05}.
In this study, however, we adopt the criterion of minimizing the
exponential moment of the codeword length $\ell(X)$, i.e.,
$\Ex[\exp\{\lambda\ell(X)\}]$ for a given parameter $\lambda>0$.

Our first contribution is to give non-asymptotic upper and lower bounds
on the exponential moment $\Ex[\exp\{\lambda\ell(X)\}]$ of the codeword
length of $\varepsilon$-source codes.  Our one-shot coding theorems
(Theorems \ref{thm:main_direct} and \ref{thm:main_converse})
demonstrate that the optimal exponential moment of the codeword length
is characterized by the smooth R\'enyi entropy.

Our second contribution is a \emph{general formula} (in the sense of
Verd\'u-Han \cite{VerduHan94,Han-spectrum}) for the asymptotic
exponential rate of the exponential moment of the codeword length
(Theorem \ref{thm:gen}).  Moreover, to apply our general formula to the
mixture of i.i.d.~sources, we analyze the asymptotic behavior of the
smooth R\'enyi entropy of the mixture of i.i.d.~sources (Theorem
\ref{thm:mixture}).

\subsection{Related Work}
The smooth R\'enyi entropy was first introduced by Renner and Wolf
\cite{RennerWolf_isit04,RennerWolf05}.  In our analysis, we use the
result of Koga \cite{Koga_itw13}, where the smooth R\'enyi entropy is
investigated by using majorization theory.  As mentioned above, the
smooth max and min R\'enyi entropies have clear operational meaning
respectively in the fixed-length source coding
\cite{RennerWolf_isit04,RennerWolf05,Uyematsu10} and the intrinsic
randomness problem
\cite{RennerWolf_isit04,RennerWolf05,UyematsuKunimatsu_itw13}.  Recently
it was shown that the smooth max R\'enyi entropy has an application also
in variable-length lossless source coding \cite{SaitoMatsushima_sita15},
where it is shown that the smooth max R\'enyi entropy characterizes the
threshold of codeword length under the condition that the overflow
probability is at most $\varepsilon$. 
Similarly, the
\emph{smooth R\'enyi divergence} also finds applications in several
coding problems; see, e.g.,
\cite{DattaRenner09,WangColbeckRenner_isit09,Warsi_itw13}.

On the other hand, conventional R\'enyi entropy \cite{Renyi61} also
plays an important role in analyses of variable-length source coding
\cite{Campbell65,Jelinek68} and fixed-length coding
\cite{Shimokawa_isit06}.  In particular, Campbell \cite{Campbell65}
proposed the exponential moment of the codeword length as an alternative
to the average codeword length as a criterion for variable-length
lossless source coding, and gave upper and lower bounds on the
exponential moment in terms of conventional R\'enyi entropy.  Our
one-shot coding theorems (Theorems \ref{thm:main_direct} and
\ref{thm:main_converse}) can be considered as a generalization of
Campbell's result to the case where the decoding error is allowed.
It should be mentioned here that a general problem for
the optimization of the exponential moment of a given cost function was
investigated by Merhav \cite{Merhav11,Merhav_arXiv2011}.

Although we consider variable-length codes subject to prefix constraints
in this paper, studies on variable-length codes \emph{without} prefix
constraints are also important
\cite{KontoyiannisVerdu14,CourtadeVerdu_isit2014}.  In particular,
Courtade and Verd\'u \cite{CourtadeVerdu_isit2014} gave non-asymptotic
upper and lower bounds on the distribution of codeword length by
bounding the cumulant generating function of the optimum codeword
lengths.  It should be noted that in \cite{KontoyiannisVerdu14} and
\cite{CourtadeVerdu_isit2014} codes are required to be injective so that
the decoder can losslessly recover the source output from the codeword.  The
problem of variable-length source coding \emph{allowing errors} was
investigated under the criterion of the average codeword length by Koga
and Yamamoto \cite{KogaYamamoto05} and Kostina \emph{et al.}
\cite{KostinaPolyanskiyVerdu14isit14,KostinaPolyanskiyVerdu14}.

\subsection{Paper Organization}
The rest of the paper is organized as follows. At first, we review the
definition of the smooth R\'enyi entropy in Section
\ref{sec:renyi_entropy}.  Then, in Section \ref{sec:theorem},
non-asymptotic coding theorems for $\varepsilon$-variable-length source
coding is given.  The general formula for the optimal exponential moment
of the codeword length achievable by $\varepsilon$-variable-length
source codes is given in Section \ref{sec:gen}.  Section
\ref{sec:conclusion} concludes the paper.  To ensure that the main ideas
are seamlessly communicated in the main text, we relegate all proofs to
the appendices.

\section{Smooth R\'enyi Entropy}\label{sec:renyi_entropy}
Renner and Wolf \cite{RennerWolf05} defined the smooth R\'enyi entropy as follows.
Fix $\varepsilon\in[0,1)$.  Given a distribution $P$ on a finite or
countably infinite set $\cX$, let $\cB^\varepsilon(P)$ be the set of
non-negative functions $Q$ with domain $\cX$ such that $Q(x)\leq P(x)$,
for all $x\in\cX$, and $\sum_{x\in\cX}Q(x)\geq 1-\varepsilon$.  Then,
for $\alpha\in(0,1)\cup(1,\infty)$, the \emph{$\varepsilon$-smooth
R\'enyi entropy of order $\alpha$} is defined as\footnote{Throughout this paper, log denotes the natural logarithm.}
\begin{align}
 H_\alpha^\varepsilon(P)\eqtri\frac{1}{1-\alpha}\log r_\alpha^\varepsilon(P)
\end{align}
where
\begin{align}
r_\alpha^\varepsilon(P)\eqtri
 \inf_{Q\in\cB^\varepsilon(P)}\sum_{x\in\cX}[Q(x)]^\alpha.
\end{align}
For basic properties of $H_\alpha^\varepsilon(P)$,  see \cite{RennerWolf05} and \cite{Koga_itw13}.

\begin{remark}
The definition of $H_\alpha^\varepsilon(P)$ above is slightly different from the original definition
given in \cite{RennerWolf_isit04}. However, in \cite{RennerWolf05}, it
is pointed out that this version is more appropriate for generalization
to conditional smooth R\'enyi entropy.  Our result in this paper
demonstrates that this version is appropriate also for describing the variable-length
source coding theorem allowing errors.
\end{remark}

\begin{remark}
 The max and min smooth R\'enyi entropies are defined respectively as
\begin{align}
 H_0^\varepsilon(P)&\eqtri \lim_{\alpha\downarrow 0}H_\alpha^\varepsilon(P),\\
 H_\infty^\varepsilon(P)&\eqtri \lim_{\alpha\to\infty}H_\alpha^\varepsilon(P).
\end{align}
As shown in \cite{RennerWolf_isit04}, $H_\alpha^\varepsilon(P)$ for $\alpha\in(0,1)$ is, up to an additive constant, equal to $H_0^\varepsilon(P)$.
This fact may be one of the reasons that $H_\alpha^\varepsilon(P)$ has received less attentions.
However, as shown in Theorems \ref{thm:main_direct}  and \ref{thm:main_converse} below, 
$H_\alpha^\varepsilon(P)$ itself plays an important role in the evaluation of the 
exponential moment of the length function. 
\end{remark}

\section{One-Shot Coding Theorem}\label{sec:theorem}
Let $\cX$ be a finite or countably infinite set and $X$ be a random variable on $\cX$ with the distribution $P$.
Without loss of generality, we assume $P(X)>0$ for all $x\in\cX$.

A variable-length source code $\code=(\varphi,\psi,\cC)$ is determined a triplet of a set $\cC\subset\{0,1\}^*$ of finite-length binary strings, an encoder mapping $\varphi\colon\cX\to\cC$, and a decoder mapping $\psi\colon\cC\to\cX$.
Without loss of generality, we assume that $\cC=\{\varphi(x):x\in\cX\}$. Further, we assume that $\cC$ satisfies the prefix condition.
The error probability of the code $\code$ is defined as
\begin{align}
 \error(\code)\eqtri\Pr\left\{X\neq\psi(\varphi(X))\right\}.
\end{align}
The length of the codeword $\varphi(x)$ of $x$ (in bits) is denoted by $\funcabs{\varphi(x)}$. Let $\ell$ be the length function (in nats):
\begin{align}
 \ell(x)\eqtri\funcabs{\varphi(x)}\log 2.
\end{align}

In this study, we focus on the exponential moment of the length function. 
For a given $\lambda>0$, let us consider the problem of minimizing
\begin{align}
 \Ex_P\left[\exp\{\lambda\ell(X)\}\right]
\end{align}
subject to $\error(\code)\leq\varepsilon$, where $\Ex_P$ denotes the expectation with respect to the distribution $P$.

\begin{remark}
In Theorems \ref{thm:main_direct} and \ref{thm:main_converse} below, we allow the encoder mapping $\varphi$ to be stochastic. 
Let $W_\varphi(c|x)$ be the probability that $x\in\cX$ is encoded in $c\in\cC$. Then,
$\error(\code)$ and $\Ex_P\left[\exp\{\lambda\ell(X)\}\right]$ are precisely written as 
\begin{align}
 \error(\code)=\sum_{x\in\cX}P(x)\sum_{c:x\neq\psi(c)}W_\varphi(c|x)
\end{align}
and
\begin{align}
\lefteqn{\Ex_P\left[\exp\{\lambda\ell(X)\}\right]}\nonumber\\
&=\sum_{x\in\cX}P(x)\sum_{c\in\cC}W_\varphi(c|x)\exp\{\lambda\funcabs{c}\log 2\}
\end{align}
where $\funcabs{c}$ is the length (in bits) of $c\in\cC$.
Note that, without loss of optimality we can assume that the decoder mapping $\psi$ is deterministic. Indeed, for a given $W_\varphi$, we can choose $\psi$ so that
\begin{align}
 \psi(c)=\arg\max W_\varphi(c|x)P(x).
\end{align}
\end{remark}

The following theorems demonstrate that the exponential moment $\Ex_P\left[\exp\{\lambda\ell(X)\}\right]$ is characterized by the smooth R\'enyi entropy $H_{1/(1+\lambda)}^\varepsilon(P)$.

\begin{theorem}
\label{thm:main_direct} 
For any $\lambda>0$ and $\varepsilon\in[0,1)$, there exists a code $\code$ (with a stochastic encoder) such that $\error(\code)\leq\varepsilon$ and
\begin{align}
\Ex_P\left[\exp\{\lambda\ell(X)\}\right]\leq 
2^{2\lambda}\exp\left\{\lambda H_{1/(1+\lambda)}^\varepsilon(P)\right\}+\varepsilon 2^{\lambda}.
\label{eq:thm_main_direct}
\end{align}
\end{theorem}

\begin{theorem}
\label{thm:main_converse} 
Fix $\lambda>0$ and $\varepsilon\in[0,1)$. Then, for any code $\code$ such that $\error(\code)\leq\varepsilon$, we have
\begin{align}
\Ex_P\left[\exp\{\lambda\ell(X)\}\right]\geq 
\exp\left\{\lambda H_{1/(1+\lambda)}^\varepsilon(P)\right\}.
\label{eq:thm_main_converse}
\end{align}
\end{theorem}

Theorem \ref{thm:main_direct} and Theorem \ref{thm:main_converse} will be proved in respectively 
Appendix \ref{sec:proof_direct} and Appendix \ref{sec:proof_converse}.

\medskip
In Theorem \ref{thm:main_direct}, we allow the encoder mapping $\varphi$
to be stochastic.  However, it is not hard to modify the theorem for the
case where only deterministic encoder mappings are allowed.
To see this, let $\cX=\{1,2,3,\dots\}$ and assume that $P(1)\geq P(2)\geq \cdots$.
Then, let $k^*=k^*(\varepsilon)$ be the minimum integer such that $\sum_{i=1}^{k^*}P(i)\geq 1-\varepsilon$
and let
\begin{align}
 Q^*(i)\eqtri
\begin{cases}
 P(i), & i=1,2,\dots, k^*-1,\\
 1-\varepsilon-\sum_{i=1}^{k^*-1}P(i), & i=k^*,\\
 0, & i> k^*.
\end{cases}
\end{align}
Since $0<1/(1+\lambda)<1$ for all $\lambda>0$, we can use (A) of Theorem 1 of \cite{Koga_itw13} and obtain
\begin{align}
 \lambda H_{1/(1+\lambda)}^\varepsilon(P)=(1+\lambda)\log\left(\sum_{i\in\cX}[Q^*(i)]^{1/(1+\lambda)}\right).
\end{align}
Based on this fact, we can modify the proof of Theorem \ref{thm:main_direct} and obtain the following result (See Appendix \ref{sec:proof_prop_deterministic} for details).

\begin{proposition}
\label{prop:deterministic}
For any $\lambda>0$ and $\varepsilon\in[0,1)$, there exists a code $\code$ with a deterministic encoder mapping $\varphi$ such that $\error(\code)\leq\varepsilon+\gamma_\varepsilon$ and
\begin{align}
\lefteqn{\Ex_P\left[\exp\{\lambda\ell(X)\}\right]}\nonumber\\
&\leq 2^{2\lambda}\exp\left\{\lambda H_{1/(1+\lambda)}^{\varepsilon+\gamma_\varepsilon}(P)\right\}+(\varepsilon+\gamma_\varepsilon) 2^{\lambda}
\end{align}
where $\gamma_\varepsilon\eqtri 1-\varepsilon-\sum_{i=1}^{k^*(\varepsilon)-1}P(i)$.
\end{proposition}

\section{General Formula}\label{sec:gen}
In this section, we consider coding problem for general sources.
A \emph{general source} 
\begin{align}
 \bX=\{X^n=(X_1^{(n)},X_2^{(n)},\dots,X_n^{(n)})\}_{n=1}^\infty
\end{align}
is defined as a sequence of random variables $X^n$ on the $n$-th Cartesian product $\cX^n$ of $\cX$ \cite{Han-spectrum}.
The distribution of $X^n$ is denoted by $P_{X^n}$, which is not required to satisfy the consistency condition.

We consider a sequence of coding problems indexed by the blocklength $n$.
A code of block length $n$ is denoted by $\code_n=(\varphi_n,\psi_n,\cC_n)$.
The length function of $\code_n$ is denoted by $\ell_n$, i.e., $\ell_n(x^n)\eqtri\funcabs{\varphi_n(x^n)}\log 2$ for all $x^n\in\cX^n$.
We are interested in the asymptotic behavior of $(1/n)\log\Ex_{P_{X^n}}[\exp\{\lambda\ell_n(X^n)\}]$.

A value $E$ is said to be $\varepsilon$-achievable if there exists a sequence  $\{\code_n\}_{n=1}^\infty$ of codes satisfying
\begin{align}
 \limsup_{n\to\infty}\error(\code_n)\leq\varepsilon
\end{align}
and
\begin{align}
 \limsup_{n\to\infty}\frac{1}{n}\log\Ex_{P_{X^n}}\left[\exp\{\lambda\ell_n(X^n)\}\right]\leq E.
\end{align}
The infimum of $\varepsilon$-achievable values is denoted by $E_\lambda^\varepsilon(\bX)$.

To characterize $E_\lambda^\varepsilon(\bX)$, we introduce the following notation.
\begin{align}
 H_{\alpha}^\varepsilon(\bX)\eqtri\lim_{\delta\downarrow 0}\limsup_{n\to\infty}\frac{1}{n}H_{\alpha}^{\varepsilon+\delta}(P_{X^n}).
\end{align}
It is worth to note that $H_{\alpha}^\varepsilon(\bX)$ is non-negative for all 
$\alpha\in(0,1)$ and $\varepsilon\in[0,1)$.
Indeed, we can prove the stronger fact that
\begin{align}
 \liminf_{n\to\infty}\frac{1}{n}H_{\alpha}^{\varepsilon}(P_{X^n})\geq 0,\quad \alpha\in(0,1),\varepsilon\in[0,1).
\label{eq:nonnegativity}
\end{align}
We will prove \eqref{eq:nonnegativity} in Appendix \ref{sec:proof_eq_nonnegativity}.

Now, we state our general formula, which will be proved in Appendix \ref{sec:proof_thm_gen}.

\begin{theorem}
\label{thm:gen}
For any $\lambda>0$ and $\varepsilon\in[0,1)$,
\begin{align}
 E_\lambda^\varepsilon(\bX)=\lambda H_{1/(1+\lambda)}^{\varepsilon}(\bX).
\end{align}
\end{theorem}

In the following, we consider a mixture of i.i.d.~sources.
Let us consider $m$ distributions 
$P_{X_i}$ ($i=1,2,\dots,m$) on $\cX$. 
A general source $\bX$ is said to be a mixture of $P_{X_1},P_{X_2},\dots,P_{X_m}$ if 
there exists $(\alpha_1,\alpha_2\dots,\alpha_m)$ satisfying $\sum_i\alpha_i=1$, $\alpha_i>0$ ($i=1,\dots,m$), and for all $n=1,2,\dots$ and all $x^n=(x_1,x_2,\dots,x_n)\in\cX^n$
\begin{align}
 P_{X^n}(x^n)&=\sum_{i=1}^m\alpha_iP_{X_i^n}(x^n)\\
&=\sum_{i=1}^m\alpha_i\prod_{t=1}^nP_{X_i}(x_t).
\label{eq:mix_iid}
\end{align}
For the later use, let $A_i\eqtri\sum_{j=1}^{i-1}\alpha_i$ ($i=1,2,\dots,m$) and $A_{m+1}\eqtri 1$.
Further, to simplify the analysis, we assume that
\begin{align}
 H(X_1)> H(X_2)>\dots>  H(X_m)
\label{eq:common_sort}
\end{align}
where
$H(X_i)$ is the entropy determined by $P_{X_i}$:
\begin{align}
 H(X_i)\eqtri\sum_{x\in\cX}P_{X_i}(x)\log\frac{1}{P_{X_i}(x)}.
\end{align}
Then, $H_{\alpha}^{\varepsilon}(\bX)$ of the mixture $\bX$ is characterized as in the following theorem.

\begin{theorem}
\label{thm:mixture}  
Let $\bX$ be a mixture of i.i.d.~sources satisfying \eqref{eq:common_sort}.
Fix $\alpha\in(0,1)$, $i$, and $\varepsilon\in[A_i,A_{i+1})$. Then, we have 
\begin{align}
 H_{\alpha}^{\varepsilon}(\bX) =H(X_i).
\label{eq:thm_mixture}
\end{align}
\end{theorem}

Theorem \ref{thm:mixture} will be proved in Appendix \ref{sec:proof_thm_mixture}.

\begin{remark}
Letting $m=1$ and $\varepsilon\downarrow 0$, Theorem \eqref{thm:mixture} derives 
Lemma I.2 of \cite{RennerWolf_isit04}. 
\end{remark}

\begin{remark}
Although Theorem \ref{thm:mixture} assumes that components are i.i.d., this assumption is not crucial.
Indeed, the property of i.i.d.~sources used in the proof of the theorem is only that the AEP \cite{Cover2} holds, i.e.,
\begin{align}
 \lim_{n\to\infty}\Pr\left\{
 \abs{
 \frac{1}{n}\log\frac{1}{P_{X_i^n}(X_i^n)}-H(X_i)} >\gamma
 \right\}=0
\end{align}
for all $i=1,2,\dots,m$ and any $\gamma>0$. 
Hence, it is straightforward to extend the theorem so that it can be applied for the mixture of stationary and ergodic sources.
Moreover, since we use only the AEP, it can be seen that the assumption
\eqref{eq:common_sort} is also not crucial.
Assume that there exists some components $j_1\neq j_2$ such as $H(X_{j_1})=H(X_{j_2})$.
Then,  let us consider the modified mixture such that ``$j_2$th component is substituted by $j_1$th component'': i.e.,
\begin{align}
 P_{X^n}(x^n)=\sum_{i\neq j_2}\alpha_i^\prime P_{X_i^n}(x^n)
\end{align}
where $\alpha_i^\prime=\alpha_i$ for $i\neq j_1$ and $\alpha_{j_1}^\prime=\alpha_{j_1}+\alpha_{j_2}$.
Then $H_{\alpha}^{\varepsilon}(\bX)$ the modified mixture is identical with the original one.
\end{remark}

Combining Theorems \ref{thm:gen} and \ref{thm:mixture}, we have the coding theorem for the mixture of i.i.d.~sources.

\begin{corollary}
Let $\bX$ be a mixture of i.i.d.\ sources satisfying \eqref{eq:common_sort}. Then, for any $\lambda>0$ and $\varepsilon\in[0,1)$,
\begin{align}
  E_\lambda^\varepsilon(\bX)=\lambda H(X_i)
\end{align}
where $i$ is determined so that $\varepsilon\in[A_i,A_{i+1})$.
\end{corollary}

\section{Concluding Remarks}\label{sec:conclusion}
In this paper, we investigated the the exponential moment of the
codeword length of variable-length source coding allowing decoding
errors.  Roughly speaking, our results demonstrate that the logarithm
$\log\Ex_P\left[\exp\{\lambda\ell(X)\}\right]$ of the optimal
exponential moment $\Ex_P\left[\exp\{\lambda\ell(X)\}\right]$ is
characterized by the smooth R\'enyi entropy
$H_{1/(1+\lambda)}^\varepsilon$.

Now, let us consider to take $\lambda\to\infty$.  When $\lambda$ is
sufficiently large, the value
$\log\Ex_P\left[\exp\{\lambda\ell(X)\}\right]$ is dominated by the
longest codeword length $\max_{x\in\cX}\ell(x)$.  In other words, to
minimize $\log\Ex_P\left[\exp\{\lambda\ell(X)\}\right]$, we need to
minimize the longest codeword length $\max_{x\in\cX}\ell(x)$.  Therefore,
roughly speaking, the difference between variable-length coding and
fixed-length coding becomes smaller as $\lambda$ is increased.  On the
other hand, we know that
$H_0^\varepsilon=\lim_{\lambda\to\infty}H_{1/(1+\lambda)}^\varepsilon$
characterizes the optimal coding rate of fixed-length codes
\cite{RennerWolf_isit04, Uyematsu10}.
The above argument implies that we can unify our result and results of \cite{RennerWolf_isit04, Uyematsu10} in the limit of $\lambda\to\infty$ or equivalently $\alpha\to 0$.

On the other hand, since $\lambda>0$ and thus $0<1/(1+\lambda)<1$, only the smooth R\'enyi entropy $H_\alpha^\varepsilon$ of the order $\alpha\in(0,1)$ plays an important role in our coding theorems.
It remains as a future work to investigate the operational meaning of the smooth R\'enyi entropy $H_\alpha^\varepsilon$ of the order $\alpha>1$.

\appendices
\section{Proof of Theorem \ref{thm:main_direct}}\label{sec:proof_direct}

Fix $\delta>0$ arbitrarily and choose $Q\in\cB^\varepsilon(P)$ so that
\begin{align}
 \log\sum_{x\in\cX}[Q(x)]^{1/(1+\lambda)}\leq \frac{\lambda}{1+\lambda}H_{1/(1+\lambda)}^\varepsilon(P) +\delta.
\label{eq1:proof_direct}
\end{align}
Let $\cA\eqtri\{x\in\cX:Q(x)>0\}$ and
\begin{align}
 \tQl(x)=\frac{[Q(x)]^{1/(1+\lambda)}}{\sum_{x'\in\cA}[Q(x')]^{1/(1+\lambda)}}.
\end{align}
Since
\begin{align}
 \sum_{x\in\cA}2^{-\{-\log_2 \tQl(x)\}}\leq 1
\end{align}
holds, we can construct $(\hat\varphi,\hat\psi,\hat\cC)$ such that (i) $\hat\cC\eqtri\{\hat\varphi(x):x\in\cA\}$ is prefix free, (ii)
$\hat\varphi\colon\cA\to\hat\cC$ satisfies
\begin{align}
 \funcabs{\hat\varphi(x)}=\lceil-\log_2 \tQl(x)\rceil,
\end{align}
and, (iii) $\hat\varphi$ and $\hat\psi\colon\hat\cC\to\cA$ satisfy $x=\psi(\varphi(x))$ for all  $x\in\cA$.

For each $x\in\cX$, let $\gamma(x)=Q(x)/P(x)$. Note that $0\leq\gamma(x)\leq 1$ and $\gamma(x)=0$ for all $x\notin\cA$.
Since $Q\in\cB^\varepsilon(P)$, we have
\begin{align}
 \sum_{x\in\cX}P(x)\gamma(x)\geq 1-\varepsilon.
\label{eq2:proof_direct}
\end{align}

Now, we construct a stochastic encoder as follows:
\begin{align}
 \varphi(x)=
\begin{cases}
 0\circ\hat\varphi(x) & \text{with probability }\gamma(x)\\
 1 & \text{with probability }1-\gamma(x)
\end{cases}
\end{align}
where $\circ$ denotes the concatenation. 
That is, $x$ is encoded to ``0'' following $\hat\varphi(x)$ with probability $\gamma(x)$, and ``1'' with probability $1-\gamma(x)$.
We can construct the corresponding decoder $\psi$ so that $x=\psi(\varphi(x))$ for all $x\in\cX$.
The length function $\ell(x)=\funcabs{\varphi(x)}\log 2$ satisfies that, if $x$ is encoded to ``0'' following $\hat\varphi(x)$, 
\begin{align}
\ell(x)&\leq -\log \tQl(x)+2\log 2
\end{align}
and otherwise $\ell(x)=\log 2$.
Hence, we have
\begin{align}
\lefteqn{\Ex_P\left[\exp\{\lambda\ell(X)\}\right]}\nonumber\\
&\leq \sum_{x\in\cX}P(x)\gamma(x)\exp\left\{\lambda[-\log \tQl(x)+2\log 2]\right\}\nonumber\\
&\qquad+\sum_{x\in\cX}P(x)(1-\gamma(x))\exp\{\lambda\log 2\}\\
&\stackrel{\text{(a)}}{\leq} 2^{2\lambda}\sum_{x\in\cA}Q(x)\exp\left\{-\lambda\log \tQl(x)\right\}
+\varepsilon 2^{\lambda}\\
&= 2^{2\lambda}\left\{\sum_{x\in\cA}[Q(x)]^{1/(1+\lambda)}\right\}^{(1+\lambda)}
+\varepsilon 2^{\lambda}\\
&\stackrel{\text{(b)}}{\leq} 2^{2\lambda}\exp\left\{\lambda H_{1/(1+\lambda)}^\varepsilon(P)+(1+\lambda)\delta\right\}
+\varepsilon 2^{\lambda}
\end{align}
where the inequality (a) follows from \eqref{eq2:proof_direct} and (b) follows from \eqref{eq1:proof_direct}.
Since we can choose $\delta>0$ arbitrarily small, we have \eqref{eq:thm_main_direct}.
\qed

\section{Proof of Theorem \ref{thm:main_converse}}\label{sec:proof_converse}

Fix a code $\code=(\varphi,\psi,\cC)$ such that $\error(\code)\leq 1-\varepsilon$.

Recall that we allow $\varphi$ to be stochastic. Let $W_\varphi(c|x)$ be the probability such that $x\in\cX$ is mapped to $c\in\cC$.
Let
\begin{align}
 \Gamma(x)\eqtri\left\{c\in\cC:W_\varphi(c|x)>0,x=\psi(c)\right\}
\end{align}
and
\begin{align}
 \gamma(x)\eqtri\sum_{c\in\Gamma(x)}W_\varphi(c|x).
\end{align}
Note that, since $\error(\code)\leq \varepsilon$, we have
\begin{align}
 \sum_{x\in\cX}P(x)\gamma(x)\geq 1-\varepsilon.
\label{eq3:converse}
\end{align}

Further, we have
\begin{align}
\lefteqn{
\Ex_P\left[\exp\{\lambda\ell(X)\}\right]}\nonumber\\
&=\sum_{x\in\cX}P(x)\sum_{c\in\cC}W_\varphi(c|x)\exp\{\lambda\funcabs{c}\log 2\}\\
&\geq \sum_{x\in\cX}P(x)\sum_{c\in\Gamma(x)}W_\varphi(c|x)\exp\{\lambda\funcabs{c}\log 2\}.
\label{eq1:converse}
\end{align}
From Jensen's inequality, it is not hard to see that
\begin{align}
\lefteqn{\sum_{c\in\Gamma(x)}
W_\varphi(c|x)\exp\{\lambda\funcabs{c}\log 2\}}\nonumber\\
&\geq \gamma(x)\exp\left\{\lambda\sum_{c\in\Gamma(x)}\frac{W_{\varphi}(c|x)}{\gamma(x)}\funcabs{c}\log 2\right\}\\
&\geq \gamma(x)\exp\left\{\lambda\sum_{c\in\Gamma(x)}\frac{W_{\varphi}(c|x)}{\gamma(x)}\bar\ell(x)\right\}\\
&= \gamma(x)\exp\left\{\lambda\bar\ell(x)\right\}
\label{eq2:converse}
\end{align}
where
\begin{align}
 \bar\ell(x)\eqtri\min_{c\in\Gamma(x)}\funcabs{c}\log 2.
\end{align}
Substituting \eqref{eq2:converse} into \eqref{eq1:converse}, we have
\begin{align}
\Ex_P\left[\exp\{\lambda\ell(X)\}\right]\geq \sum_{x\in\cX}P(x)\gamma(x)\exp\left\{\lambda\bar\ell(x)\right\}.
\label{eq4:converse}
\end{align}
Let $Q(x)=P(x)\gamma(x)$. Then, from \eqref{eq3:converse}, we have $Q\in\cB^\varepsilon(P)$.
Let $\cA=\{x:Q(x)>0\}$. Then, \eqref{eq4:converse} can be written as
\begin{align}
\Ex_P\left[\exp\{\lambda\ell(X)\}\right]\geq \sum_{x\in\cA}Q(x)\exp\left\{\lambda\bar\ell(x)\right\}.
\label{eq6:converse}
\end{align}

On the other hand, from the definition of the set $\Gamma(x)$, we can see that
$\Gamma(x)\cap\Gamma(x')=\emptyset$ for all $x,x'\in\cA$ such that $x\neq x'$, and thus we have
\begin{align}
 \sum_{x\in\cA}\exp\{-\bar\ell(x)\}\leq 1.
\label{eq5:converse}
\end{align}

Now, let us consider the problem of minimizing
$\sum_{x\in\cA}Q(x)\exp\left\{\lambda\bar\ell(x)\right\}$ subject to \eqref{eq5:converse}.
As shown in Example 1 in Section 3 of \cite{Merhav_arXiv2011}, the minimum is achieved by
\begin{align}
 \bar\ell(x)=-\log\frac{[Q(x)]^{1/(1+\lambda)}}{\sum_{x'\in\cA}[Q(x')]^{1/(1+\lambda)}},\quad x\in\cA.
\end{align}
In other words, \eqref{eq6:converse} can be rewritten as
\begin{align}
\lefteqn{\Ex_P\left[\exp\{\lambda\ell(X)\}\right]}\nonumber\\
&\geq \sum_{x\in\cA}Q(x)\exp\left\{-\lambda
\log\frac{[Q(x)]^{1/(1+\lambda)}}{\sum_{x'\in\cA}[Q(x')]^{1/(1+\lambda)}}
\right\}\\
&=\left[\sum_{x\in\cA}[Q(x)]^{1/(1+\lambda)}\right]^{(1+\lambda)}\\
&\geq\left[r_{1/(1+\lambda)}^\varepsilon(P)\right]^{(1+\lambda)}
\end{align}
where the last inequality follows from the fact $Q\in\cB^\varepsilon(P)$.
By the definition of the smooth R\'enyi entropy, we have \eqref{eq:thm_main_converse}.
\qed

\section{Proof of Proposition \ref{prop:deterministic}}\label{sec:proof_prop_deterministic}
Let 
\begin{align}
 \hQs(i)\eqtri
\begin{cases}
 P(i), & i=1,2,\dots,k^*(\varepsilon)-1,\\
 0, & i>k^*(\varepsilon).
\end{cases}
\end{align}
Then, from Theorem 1 (A) of \cite{Koga_itw13}, we have
\begin{align}
 \lambda H_{1/(1+\lambda)}^{\varepsilon+\gamma_\varepsilon}(P)=(1+\lambda)\log\left(\sum_{i\in\cX}[Q^*(i)]^{1/(1+\lambda)}\right).
\end{align}
Now, let us substitute $\varepsilon$ (resp.~$Q$) in the proof of Theorem \ref{thm:main_direct} 
with $\varepsilon+\gamma_\varepsilon$ (resp.~$\hQs$).
Note that $\gamma(x)=\hQs(x)/P(x)$ satisfies 
\begin{align}
 \gamma(x)=
\begin{cases}
 1, & i=1,2,\dots,k^*(\varepsilon)-1,\\
 0, & i>k^*(\varepsilon).
\end{cases}
\end{align}
Thus, the encoder constructed in the proof of Theorem \ref{thm:main_direct} becomes deterministic.
Hence, we can obtain the proposition.\qed

\section{Proof of \eqref{eq:nonnegativity}}\label{sec:proof_eq_nonnegativity}
Fix $\alpha\in(0,1)$ and $\varepsilon\in[0,1)$, and then, choose $\varepsilon'>0$ so that $\varepsilon+\varepsilon'<1$.
From Lemma 2 of \cite{RennerWolf05}, we have
\begin{align}
\frac{1}{n}
H_\alpha^\varepsilon(P_{X^n})\geq 
\frac{1}{n}H_0^{\varepsilon+\varepsilon'}(P_{X^n})
-\frac{\log(1/\varepsilon')}{n(1-\alpha)}.
\label{eq:proof_eq_nonnegativity}
\end{align}
On the other hand, it is known that $H_0^{\varepsilon+\varepsilon'}(P_{X^n})$ can be written as
\begin{align}
H_0^{\varepsilon+\varepsilon'}(P_{X^n})
&= \min_{\substack{\cA\subseteq\cX\\ P(\cA)\geq 1-\varepsilon-\varepsilon'}}\log\abs{\cA},
\end{align}
where $\abs{\cA}$ is the cardinality of $\cA$, and thus, $H_0^{\varepsilon+\varepsilon'}(P_{X^n})\geq 0$.
So, taking the inferior limit of both sides of \eqref{eq:proof_eq_nonnegativity}, we have \eqref{eq:nonnegativity}.
\qed

\section{Proof of Theorem \ref{thm:gen}}\label{sec:proof_thm_gen}
\begin{proof}
[Direct Part]
At first, we consider the case where
\begin{align}
 H_{1/(1+\lambda)}^{\varepsilon}(\bX)>0.
\end{align}
In this case, for all sufficiently small $\delta>0$ and sufficiently large $n$, we have
\begin{align}
 2^{2\lambda}\exp\left\{\lambda H_{1/(1+\lambda)}^{\varepsilon+\delta}(P_{X^n})\right\}>\varepsilon 2^{\lambda}.
\end{align}
Hence, from Theorem \ref{thm:main_direct}, there exists $\{\code_n\}_{n=1}^\infty$ such that
\begin{align}
 \error(\code_n)\leq\varepsilon+\delta,\quad n=1,2,\dots,
\end{align}
and, for sufficiently large $n$,
\begin{align}
\Ex_P\left[\exp\{\lambda\ell_n(X^n)\}\right]\leq 
2\times2^{2\lambda}\exp\left\{\lambda H_{1/(1+\lambda)}^{\varepsilon+\delta}(P_{X^n})\right\}.
\label{eq1:proof_prop_gen}
\end{align}
Eq.~\eqref{eq1:proof_prop_gen} gives
\begin{align}
\lefteqn{
 \limsup_{n\to\infty}\frac{1}{n}\log \Ex_P\left[\exp\{\lambda\ell_n(X^n)\}\right]}\nonumber\\
&\leq \lambda \limsup_{n\to\infty}\frac{1}{n}H_{1/(1+\lambda)}^{\varepsilon+\delta}(P_{X^n}).
\end{align}
By using the \emph{diagonal line argument} (see \cite{Han-spectrum}), we can conclude that $\lambda H_{1/(1+\lambda)}^\varepsilon(\bX)$ is $\varepsilon$-achievable.

If $H_{1/(1+\lambda)}^{\varepsilon}(\bX)=0$ then \eqref{eq1:proof_prop_gen} is replaced with
\begin{align}
\lefteqn{
 \Ex_P\left[\exp\{\lambda\ell_n(X^n)\}\right]}\nonumber\\&\leq 
\max\left\{
2\times2^{2\lambda}\exp\left\{\lambda H_{1/(1+\lambda)}^{\varepsilon+\delta}(P_{X^n})\right\},
2\times\varepsilon 2^{\lambda}
\right\}
\end{align}
In this case, we can also prove that $0$ is $\varepsilon$-achievable in the same way as the case $H_{1/(1+\lambda)}^\varepsilon(\bX)>0$.
\end{proof}

\begin{proof}
[Converse Part]
Suppose that $E$ is $\varepsilon$-achievable and fix $\delta>0$ arbitrarily.
Then there exists $\{\code_n\}_{n=1}^\infty$ such that, for sufficiently large $n$,
\begin{align}
 \error(\code_n)\leq \varepsilon+\delta
\label{eq4:proof_prop_gen}
\end{align}
and
\begin{align}
 \limsup_{n\to\infty}\frac{1}{n}\log \Ex_P\left[\exp\{\lambda\ell_n(X^n)\}\right]\leq E.
\label{eq2:proof_prop_gen}
\end{align}
On the other hand, from Theorem \ref{thm:main_converse}, for sufficiently large $n$ such that 
\eqref{eq4:proof_prop_gen}
holds, 
\begin{align}
 \Ex_P\left[\exp\{\lambda\ell_n(X^n)\}\right]\geq \exp\left\{\lambda H_{1/(1+\lambda)}^{\varepsilon+\delta}(P_{X^n})\right\}.
\label{eq3:proof_prop_gen}
\end{align}
Combining \eqref{eq2:proof_prop_gen} with \eqref{eq3:proof_prop_gen}, we have
\begin{align}
 E\geq \lambda\limsup_{n\to\infty}\frac{1}{n}H_{1/(1+\lambda)}^{\varepsilon+\delta}(P_{X^n}).
\end{align}
Since $\delta>0$ is arbitrary, letting $\delta\downarrow 0$, we have
$E\geq \lambda H_{1/(1+\lambda)}^{\varepsilon}(\bX)$.
\end{proof}

\section{Proof of Theorem \ref{thm:mixture}}\label{sec:proof_thm_mixture}
\subsection{Lemmas}
Before proving the theorem, we introduce some lemmas.

\begin{lemma}
\label{lemma:tmp1_proof_thm_common}
Fix $\gamma>0$ arbitrarily. Then, there exists an integer $n_0$ so that for all $n\geq n_0$ and all $i=1,2,\dots,m$,
\begin{align}
 \Pr\left\{\frac{1}{n}\log\frac{1}{P_{X^n}(X^n)}\geq H(X_i)-\gamma\right\}\geq A_{i+1}-\gamma.
\end{align}
\end{lemma}

\begin{IEEEproof}
For each $k=1,2,\dots,m$, let
\begin{align}
 \cS_k^n\eqtri\left\{x^n:\frac{1}{n}\log\frac{1}{P_{X_k^n}(x^n)}\geq H(X_k)-\frac{\gamma}{2}\right\}.
\end{align}
Since i.i.d.~sources satisfy the AEP \cite{Cover2}, we can choose $n_1$ such that
\begin{align}
 \sum_{x^n\in\cS_k^n}P_{X_k^n}(x^n)\geq 1-\frac{\gamma}{2},\quad\forall n\geq n_1,\forall k=1,2,\dots,m.
\end{align}
Moreover, we can choose $n_0\geq n_1$ so that
\begin{align}
 -\frac{1}{n}\log\gamma\leq\frac{\gamma}{2},\quad\forall n\geq n_0.
\end{align}

Then, for all $n\geq n_0$, any $i=1,2,\dots,m$, and any $k=1,2,\dots,i$,
\begin{align}
 \tcS_i^n\eqtri\left\{x^n:\frac{1}{n}\log\frac{1}{P_{X^n}(x^n)}\geq H(X_i)-\gamma\right\}
\end{align}
and
\begin{align}
 \cT_k^n\eqtri\left\{x^n:P_{X_k^n}(x^n)\leq\gamma P_{X^n}(x^n)\right\}
\end{align}
satisfy that
\begin{align}
 \tcS_i^n\cup\cT_k^n
&\supseteq\left\{
x^n:\frac{1}{n}\log\frac{\gamma}{P_{X_k^n}(x^n)}\geq H(X_i)-\gamma
\right\}\\
&=\left\{x^n:\frac{1}{n}\log\frac{1}{P_{X_k^n}(x^n)}\geq H(X_i)-\gamma-\frac{1}{n}\log\gamma\right\}\\
&\supseteq\left\{x^n:\frac{1}{n}\log\frac{1}{P_{X_k^n}(x^n)}\geq H(X_i)-\frac{\gamma}{2}\right\}\\
&\supseteq\cS_k^n.
\end{align}
Thus, we have
\begin{align}
 \sum_{x^n\in\tcS_i^n}P_{X^n}(x^n)
&\geq  \sum_{k=1}^i\alpha_k\sum_{x^n\in\tcS_i^n}P_{X_k^n}(x^n)\\
&\geq  \sum_{k=1}^i\alpha_k\sum_{x^n\in\cS_k^n}P_{X_k^n}(x^n)-\sum_{k=1}^i\alpha_k \sum_{x^n\in\cT_k^n}P_{X_k^n}(x^n)  \\
&\geq A_{i+1}(1-\gamma/2)-\sum_{k=1}^i\alpha_k \sum_{x^n\in\cT_k^n}\frac{\gamma}{2}P_{X^n}(x^n)  \\
&\geq A_{i+1}(1-\gamma/2)-\frac{\gamma}{2}  \\
&\geq A_{i+1}-\gamma.
\end{align}
\end{IEEEproof}

\begin{lemma}
\label{lemma:tmp2_proof_thm_common}
Fix $\gamma>0$ arbitrarily. Then, there exists an integer $n_0$ so that for all $n\geq n_0$ and all $i=1,2,\dots,m$,
\begin{align}
 \Pr\left\{\frac{1}{n}\log\frac{1}{P_{X^n}(X^n)}\leq H(X_i)+\gamma\right\}\geq 1-A_i-\gamma.
\end{align}
\end{lemma}

\begin{IEEEproof}
For each $k=1,2,\dots,m$, let
\begin{align}
 \cS_k^n\eqtri\left\{x^n:\frac{1}{n}\log\frac{1}{P_{X_k^n}(x^n)}\leq H(X_k)+\frac{\gamma}{2}\right\}.
\end{align}
Since i.i.d.~sources satisfy the AEP \cite{Cover2}, we can choose $n_1$ such that
\begin{align}
 \sum_{(x^n,y^n)\in\cS_k^n}P_{X_k^nY_k^n}(x^n,y^n)\geq 1-\gamma,\quad\forall n\geq n_1,\forall k=1,2,\dots,m.
\end{align}
Moreover, we can choose $n_0\geq n_1$ so that
\begin{align}
 -\frac{1}{n}\log\alpha_k\leq\frac{\gamma}{2},\quad\forall n\geq n_0,\forall k=1,2,\dots,m.
\end{align}

Hence, for all $n\geq n_0$ and any $i$,
\begin{align}
 \tcS_i^n\eqtri\left\{x^n:\frac{1}{n}\log\frac{1}{P_{X^n}(x^n)}\leq H(X_k)+\gamma\right\}
\end{align}
satisfies that
\begin{align}
\tcS_i^n
&\supseteq\left\{x^n:\frac{1}{n}\log\frac{1}{\alpha_k P_{X_k^n}(x^n)}\leq H(X_i)+\gamma\right\}\\
&=\left\{x^n:\frac{1}{n}\log\frac{1}{P_{X_k^n}(x^n)}\leq H(X_i)+\gamma+\frac{1}{n}\log\alpha_k\right\}\\
&\supseteq\cS_k^n.
\end{align}
Thus, we have
\begin{align}
 \sum_{x^n\in\tcS_i^n}P_{X^n}(x^n)
&\geq  \sum_{k=i}^m\alpha_k\sum_{x^n\in\tcS_i^n}P_{X_k^n}(x^n)\\
&\geq  \sum_{k=i}^m\alpha_k\sum_{x^n\in\cS_k^n}P_{X_k^n}(x^n)\\
&\geq (1-A_i)(1-\gamma)\\
&\geq 1-A_i-\gamma.
\end{align}
\end{IEEEproof}

\begin{lemma}
\label{lemma:tmp3_proof_thm_common}
Fix $\gamma>0$ so that $H(X_j)-\gamma>H(X_{j+1})+\gamma$ for all $j=1,2,\dots,m-1$.
Then, for sufficiently large $n$ and any $i=1,2,\dots,m$,
\begin{align}
 \alpha_i-2\gamma\leq \Pr\left\{\abs{\frac{1}{n}\log\frac{1}{P_{X^n}(X^n)}- H(X_i)}\leq\gamma\right\}\leq \alpha_i+2\gamma.
\end{align}
\end{lemma}

\begin{IEEEproof}
From Lemmas \ref{lemma:tmp1_proof_thm_common} and \ref{lemma:tmp2_proof_thm_common}, we have
\begin{align}
\lefteqn{
 \Pr\left\{\abs{\frac{1}{n}\log\frac{1}{P_{X^n}(X^n)}- H(X_i)}\leq\gamma\right\}
}\nonumber\\
&=  \Pr\left\{\frac{1}{n}\log\frac{1}{P_{X^n}(X^n)}\leq H(X_i)+\gamma\right\}- \Pr\left\{\frac{1}{n}\log\frac{1}{P_{X^n}(X^n)}< H(X_i)-\gamma\right\}\\
&\geq \{1-A_i-\gamma\}-\{1-(A_{i+1}-\gamma)\}\\
&=\alpha_i-2\gamma
\end{align}
and
\begin{align}
 \lefteqn{
 \Pr\left\{\abs{\frac{1}{n}\log\frac{1}{P_{X^n}(X^n)}- H(X_i)}\leq\gamma\right\}
}\nonumber\\
&=  \Pr\left\{\frac{1}{n}\log\frac{1}{P_{X^n}(X^n)}\leq H(X_i)+\gamma\right\}- \Pr\left\{\frac{1}{n}\log\frac{1}{P_{X^n}(X^n)}< H(X_i)-\gamma\right\}\\
&\leq  \Pr\left\{\frac{1}{n}\log\frac{1}{P_{X^n}(X^n)}< H(X_{i-1})-\gamma\right\}- \Pr\left\{\frac{1}{n}\log\frac{1}{P_{X^n}(X^n)}\leq H(X_{i+1})+\gamma\right\}\\
&\leq \{1-(A_i-\gamma)\}-\{1-A_{i+1}-\gamma\}\\
&=\alpha_i+2\gamma.
\end{align}
\end{IEEEproof}

\subsection{Proof of Theorem \ref{thm:mixture}}

To prove the theorem, it is sufficient to show that, for $\varepsilon$ satisfying $A_i<\varepsilon<A_{i+1}$, 
\begin{align}
 \limsup_{n\to\infty}\frac{1}{n}H_\alpha^\varepsilon(P_{X^n})&\leq H(X_i)\label{goal1:proof_thm_mixture}
\intertext{and}
 \liminf_{n\to\infty}\frac{1}{n}H_\alpha^\varepsilon(P_{X^n})&\geq H(X_i).\label{goal2:proof_thm_mixture}
\end{align}

\begin{IEEEproof}
[Proof of \eqref{goal1:proof_thm_mixture}]
Fix $\gamma>0$ sufficiently small so that $H(X_j)-\gamma>H(X_{j+1})+\gamma$ for all $j=1,2,\dots,m-1$ and that
$A_i+2m\gamma<\varepsilon$.
For $j=1,2,\dots,m$, let
\begin{align}
 \cT_n(j)\eqtri\left\{
 x^n:\abs{
 \frac{1}{n}\log\frac{1}{P_{X^n}(x^n)}-H(X_j)
 }\leq \gamma
 \right\}.
\label{eq:proof_thm_mixture_defT}
\end{align}
Note that $\cT_n(j)\cap\cT_n(\hat{j})=\emptyset$ ($j\neq\hat{j}$).
Further, from Lemma \ref{lemma:tmp3_proof_thm_common}, we have
\begin{align}
 \Pr\left\{X^n\in \bigcup_{j=i}^m\cT_n(j)\right\}
&= \sum_{j=i}^m \Pr\left\{X^n\in \cT_n(j)\right\}\\
&\geq \sum_{j=i}^m \left(\alpha_j-2\gamma\right)\\
&\geq 1-A_i-2m\gamma\\
&\geq 1-\varepsilon. \label{eq2:proof_thm_mixture}
\end{align}
From \eqref{eq2:proof_thm_mixture}, we can see that
\begin{align}
 Q_n(x^n)\eqtri
\begin{cases}
 P_{X^n}(x^n), & \text{if }x^n\in\bigcup_{j=i}^m\cT_n(j)\\
 0, & \text{otherwise}
\end{cases}
\end{align}
satisfies $Q_n\in\cB^\varepsilon(P_{X^n})$. Thus, from the definition of $r_\alpha^\varepsilon(P_{X^n})$,
\begin{align}
 r_\alpha^\varepsilon(P_{X^n})
&\leq\sum_{x^n\in\cX^n}[Q_n(x^n)]^\alpha\\
&=\sum_{j=i}^m\sum_{x^n\in\cT_n(j)}[P_{X^n}(x^n)]^\alpha\\
&\leq \sum_{j=i}^m\abs{\cT_n(j)}\exp\{-\alpha n(H(X_j)-\gamma)\}\\
&\leq \sum_{j=i}^m\exp\{n(H(X_j)+\gamma)\}\exp\{-\alpha n(H(X_j)-\gamma)\}\\
&= \sum_{j=i}^m\exp\{n[(1-\alpha)H(X_j)+(1+\alpha)\gamma]\}\\
&\leq m\exp\{n[(1-\alpha)H(X_i)+(1+\alpha)\gamma]\}.
\end{align}
Hence, we have
\begin{align}
 \frac{1}{n}H_\alpha^\varepsilon(P_{X^n})\leq H(X_i)+\frac{1+\alpha}{1-\alpha}\gamma+\frac{1}{n}\log m
\end{align}
and thus
\begin{align}
 \limsup_{n\to\infty}\frac{1}{n}H_\alpha^\varepsilon(P_{X^n})&\leq H(X_i)+\frac{1+\alpha}{1-\alpha}\gamma.
\end{align}
Since we can choose $\gamma>0$ arbitrarily small, we have \eqref{goal1:proof_thm_mixture}.
\end{IEEEproof}

\begin{IEEEproof}
[Proof of \eqref{goal2:proof_thm_mixture}]
If $H(X_i)=0$ then \eqref{goal2:proof_thm_mixture} is apparent, since \eqref{eq:nonnegativity} holds.
So, we assume $H(X_i)>0$.

Fix $\gamma>0$ sufficiently small so that $H(X_j)-\gamma>H(X_{j+1})+\gamma$ for all $j=1,2,\dots,m-1$ and that
$A_i+6m\gamma<\varepsilon<A_{i+1}-6m\gamma$.
We assume that $n$ is sufficiently large so that $\exp\{-n[H(X_i)-\gamma]\}\leq m\gamma$.
Let us define $\cT_n(j)$ as in \eqref{eq:proof_thm_mixture_defT}. 
Note that 
\begin{align}
 P_{X^n}(x^n)<P_{X^n}(\hat{x}^n),\quad x^n\in\cT_n(j), \hat{x}^n\in\cT_n(\hat{j}), j<\hat{j}.
\label{eq4:proof_goal2}
\end{align}
Let $\cS_n\eqtri\bigcup_{j=1}^{m}\cT_n(j)$ and $\bcS_n\eqtri\cX^n\setminus\cS_n$.
Then, from Lemma \ref{lemma:tmp3_proof_thm_common}, we have
\begin{align}
 P_{X^n}(\bcS_n)\leq 2m\gamma.
\label{eq1:proof_goal2}
\end{align}

Let us sort the sequences in $\cX^n$ so that
\begin{align}
 P_{X^n}(x_1^n)\geq P_{X^n}(x_2^n)\geq P_{X^n}(x_3^n)\geq\dots.
\end{align}
Then, let $\cA_n\eqtri\{x_1^n,x_2^n,\dots,x_{k^*-1}^n\}$ and $\cA_n^{+}\eqtri\cA_n\cup\{x_{k^*}^n\}$ 
where $k^*$ is the integer satisfying
\begin{align}
 \sum_{k=1}^{k^*}P_{X^n}(x_k^n)&\geq 1-\varepsilon
\intertext{and}
 \sum_{k=1}^{k^*-1}P_{X^n}(x_k^n)&< 1-\varepsilon.
\end{align}

We first show that
\begin{align}
x_{k^*}^n\in\bcS_n\text{ or }x_{k^*}^n\in\bigcup_{j=1}^i\cT_n(j).
\label{eq2:proof_goal2}
\end{align}
From Lemma \ref{lemma:tmp3_proof_thm_common}, we have
\begin{align}
 \Pr\left\{X^n\in\bigcup_{j=i+1}^m\cT_n(j)\right\}&\leq \sum_{j=i+1}^m(\alpha_j+2\gamma)\\
&\leq 1-A_{i+1}+2m\gamma\\
&\leq 1-\varepsilon-4m\gamma.
\label{eq3:proof_goal2}
\end{align}
Since $P(\cA_n^{+})\geq 1-\varepsilon$ holds, from \eqref{eq1:proof_goal2} and \eqref{eq3:proof_goal2}, we have
\begin{align}
 \cA_n^{+}\cap\left[\bigcup_{j=1}^i\cT_n(j)\right]\neq\emptyset.
\label{eq5:proof_goal2}
\end{align}
From \eqref{eq4:proof_goal2} and \eqref{eq5:proof_goal2}, we can obtain \eqref{eq2:proof_goal2}.

We next notice that, from \eqref{eq4:proof_goal2} and the assumption that $n$ is sufficiently large, we have $P_{X^n}(x^n)\leq\exp\{-n[H(X_i)-\gamma]\}\leq m\gamma$ for all $x^n\in\bigcup_{j=1}^i\cT_n(j)$.
Combining this fact with \eqref{eq1:proof_goal2} and \eqref{eq2:proof_goal2}, we can see that 
\begin{align}
 P_{X^n}(\cA_n\cap\cS_n)&\geq 1-\varepsilon-m\gamma-P_{X^n}(\bcS_n)\\
&\geq 1-\varepsilon-3m\gamma.
\label{eq6:proof_goal2}
\end{align}
Thus, from \eqref{eq3:proof_goal2} and \eqref{eq6:proof_goal2}, we have
\begin{align}
\Pr\left\{X^n\in \cA_n\cap\left[\bigcup_{j=1}^i\cT_n(j)\right]\right\}\geq m\gamma.
\label{eq7:proof_goal2}
\end{align}
Moreover, since \eqref{eq4:proof_goal2} holds, \eqref{eq7:proof_goal2} implies that
\begin{align}
P_{X^n}(\cA_n\cap\cT_n(i))\geq \beta\eqtri\min\{m\gamma,\alpha_i\}
\end{align}
and thus
\begin{align}
\abs{\cA_n\cap\cT_n(i)}\geq \beta\exp\{n[H(X_i)-\gamma]\}.
\end{align}
Hence, we have
\begin{align}
\sum_{x^n\in\cA_n\cap\cT_n(i)}[P_{X^n}(x^n)]^\alpha\geq \beta\exp\{n[(1-\alpha)H(X_i)-(1+\alpha)\gamma]\}.
\label{eq8:proof_goal2}  
\end{align}

Now we use the result of Koga \cite{Koga_itw13}.
Theorem 1 (A) of \cite{Koga_itw13} tells us that
\begin{align}
 H_\alpha^\varepsilon(P_{X^n})&\geq\frac{1}{1-\alpha}\log\left(\sum_{k=1}^{k^*-1}[P_{X^n}(x_k^n)]^\alpha\right)\\
&=\frac{1}{1-\alpha}\log\left(\sum_{x^n\in\cA_n}[P_{X^n}(x^n)]^\alpha\right).
\end{align}
By combining this with \eqref{eq8:proof_goal2}, we have
\begin{align}
\frac{1}{n}H_\alpha^\varepsilon(P_{X^n})&\geq\frac{1}{n(1-\alpha)}\log\left(\sum_{x^n\in\cA_n\cap\cT_n(i)}[P_{X^n}(x^n)]^\alpha\right)\\
&\geq\frac{1}{n(1-\alpha)}\log\left(\beta\exp\{n[(1-\alpha)H(X_i)-(1+\alpha)\gamma]\}\right)\\
&=H(X_i)-\frac{1+\alpha}{1-\alpha}\gamma+\frac{\log\beta}{n(1-\alpha)}.
\end{align}
Thus, we have
\begin{align}
 \liminf_{n\to\infty}\frac{1}{n}H_\alpha^\varepsilon(P_{X^n})\geq H(X_i)-\frac{1+\alpha}{1-\alpha}\gamma.
\end{align}
Since we can choose $\gamma>0$ arbitrarily small, we have \eqref{goal2:proof_thm_mixture}.
\end{IEEEproof}

\section*{Acknowledgment}
The author would like to thank Prof.~Mitsugu Iwamoto for informing him about \cite{Campbell65}.

\bibliographystyle{IEEETran}
\bibliography{reference}

\end{document}